\documentclass[pra, aps, twocolumn, floatfix, superscriptaddress, amsmath,  longbibliography]{revtex4-2}
\usepackage{graphicx}
\usepackage{graphics}
\usepackage{mathtools}
\usepackage{float}
\usepackage{amssymb}
\usepackage{bbold}

\usepackage{siunitx}

\usepackage[dvipsnames]{xcolor}
\definecolor{darkblue}{rgb}{0.0, 0.0, 0.75}

\usepackage{color}
\usepackage[colorlinks=true,
            linkcolor=darkblue,
            urlcolor=darkblue,
            citecolor=darkblue]{hyperref}

	\usepackage[normalem]{ulem} 
	\definecolor{mgreen}{RGB}{1,123,0}

\def \br{{\bf r}}

\def \mb{\text{b}}
\def \mh{\text{h}}

\def \mref{\mathrm{ref}}

\def \mB{\mathrm{B}}

\def \mring{\mathrm{ring}}

\def \mVR{\mathrm{VR}}

\def \mRP{\mathrm{RP} }

\def \mHz{\mathrm{Hz}}

\def \mms{\mathrm{ms}}

\begin{document}
%
\title{Controlled generation of 3D vortices in driven atomic Josephson junctions}
%
\author{Vijay Pal Singh}
\affiliation{Quantum Research Centre, Technology Innovation Institute, Abu Dhabi, UAE}
\author{Ludwig Mathey}
\affiliation{Zentrum f\"ur Optische Quantentechnologien and Institut f\"ur  Quantenphysik, Universit\"at Hamburg, 22761 Hamburg, Germany}
\affiliation{The Hamburg Centre for Ultrafast Imaging, Luruper Chaussee 149, Hamburg 22761, Germany}
\author{Herwig Ott}
\affiliation{Department of Physics and Research Center OPTIMAS, Rheinland-Pf\"alzische Technische Universit\"at Kaiserslautern-Landau, 67663 Kaiserslautern, Germany}
\author{Luigi Amico}
\affiliation{Quantum Research Centre, Technology Innovation Institute, Abu Dhabi, UAE}
\affiliation{Dipartimento di Fisica e Astronomia, Universit\`a di Catania, Via S. Sofia 64, 95123 Catania, Italy}
\affiliation{INFN-Sezione di Catania, Via S. Sofia 64, 95127 Catania, Italy}

\date{\today}
%

%
\begin{abstract}
We propose an ac-driven atomic Josephson junction as a clean and tunable source of three dimensional (3D) solitary waves in quantum fluids.  
Depending on the height of the junction barrier, the emitted excitations appear as vortex rings at low velocity or vorticity-free rarefaction pulses near the sound velocity, thus spanning the complete Jones-Roberts family of solitons.
The Shapiro-step phenomenon renders the emission deterministic: on the first, second, third Shapiro steps, the junction ejects one, two, and three solitary excitations per drive cycle. 
This enables controlled generation of single- and multi-excitation configurations, allowing detailed studies of the full crossover between vortex rings and rarefaction pulses and their interaction dynamics. 
In particular, deterministic multi-ring emission provides insights into  leapfrogging dynamics of two and three coaxial rings and their decay via boundary-assisted, sound-mediated processes.   
This ac-driven protocol establishes a compact and reproducible platform for generating, classifying, and controlling 3D solitonic excitations, paving the way for precision studies of nonlinear vortex dynamics, dissipation, and quantum turbulence in trapped superfluids.  
\end{abstract}
\maketitle
%
%

Quantum vortex dynamics lies at the core of diverse physical phenomena, ranging  from turbulence  and astrophysical plasmas to emerging quantum technologies \cite{Barenghi2001, vinen2002quantum,Tsubota2013}. 
Superfluid helium \cite{Bradley2006,Walmsley2008} and, more recently, ultracold atomic gases \cite{Navon2016,Navon2019,Dogra2023,Adams1999, Bradley2012, Neely2013, Kwon2014, Karailiev2024,fetter2001vortices} provide powerful and highly controllable platforms to explore these quantized vortices under well-defined conditions. These systems enable fundamental questions, such as vortex-vortex interactions, stability, and the transition from classical to quantum turbulence, to be investigated with unprecedented accuracy \cite{tsubota2010quantized}.
In two-dimensional (2D) condensates, major insights have been achieved through the controlled generation and manipulation of vortex-antivortex pairs using phase imprinting \cite{neely2010observation,aioi2011controlled,samson2016deterministic} or the so-called `chopsticks' method \cite{kwon2021sound}. In non-simply connected geometries, vortex dynamics sustains quantized persistent currents \cite{polo2025persistent}, underpinning key developments in the emerging field of atomtronics \cite{Amico2021,Amico2022}. 
The dynamics of 3D vortices, however, poses an additional level of complexity. Dissipation \cite{vinen2005turbulent}, the interplay between vortex tangles and lines \cite{nemirovskii2014reconnection}, and reconnections \cite{alamri2008reconnection,bewley2008characterization,paoletti2010reconnection} represent 
formidable challenges for both real-time numerical simulations and experiments \cite{barenghi2014introduction,barenghi2014experimental,guthrie2021nanoscale,nemirovskii2013quantum,Ruostekoski2001,Schloss2020}. 
Establishing a controlled platform to generate and analyze 3D vortex dynamics is therefore of great importance.

A cornerstone in this context is the family of Jones-Roberts solitary-wave solutions of the 3D Gross-Pitaevskii equation (GPE).
Using a combination of numerical and analytical methods, Jones and Roberts identified a continuous branch of solutions that smoothly connects quantized vortex rings (VRs) to vorticity-free rarefaction pulses (RPs) \cite{Jones1982, Jones1986}. 
In real dynamics, these excitations manifest as transient structures that decay through sound emission, vortex reconnections, or Kelvin-wave cascades. 
The low-velocity branch corresponds to VRs carrying quantized circulation.  
As the momentum decreases, the ring velocity increases up to a finite energy-momentum cusp, 
beyond which the branch connects continuously to RPs propagating near the sound velocity. 
This unified family thus reveals a natural dynamical pathway by which a VR can shrink, shed its vorticity at the cusp, and transform into an RP-like excitation.

While specific features of these solitonic excitations have been explored theoretically \cite{Jones1986,Feder2000, Brand2002,Theocharis2003,Mamaev1996, Becker2013, Xhani2020, Xhani2020NJP} and experimentally 
\cite{Anderson2001, Ginsberg2005, Shomroni2009, Burchianti2018}, 
a systematic experimental probe of the full JR family---from vortex rings to rarefaction pulses---has remained elusive. 
Previous theoretical studies have shown that VRs interact via Biot-Savart-like induction:  
generic, non-coaxial collisions excite Kelvin waves along the filaments, 
leading to sound emission and eventual ring breakup 
\cite{Berloff2004,Caplan2014,Ikuta2019,Pan2022,Leadbeater2003, White2010, Bai2025}.
For coaxial configurations, vortex rings exhibit `leapfrogging' dynamics, 
alternately decelerating and accelerating as they exchange momentum.

 Experiments in superfluid helium have visualized rings, reconnections, and sound-mediated decay, 
thereby anchoring these mechanisms experimentally  \cite{Hanninen2007, Tang2023}.
In trapped atomic condensates, finite temperature and spatial inhomogeneity introduce additional damping 
(mutual friction) that accelerates shrinkage and decay \cite{Barenghi2009,Reichl2013, SinghAJJ2025}.

 Here, we propose driven atomic Josephson junctions (AJJs) as a tunable and reproducible platform for generating JR excitations on demand and mapping their dynamics.  
Relying on Shapiro steps, we demonstrate that by tuning the barrier strength and drive amplitude, each drive cycle can be engineered to produce either VRs or RPs. 
We then (i) extract ring radii and velocities from 3D isosurfaces and vortex-filament reconstructions, 
(ii) anaylze the column-density deficit as a proxy for interaction energy, 
and (iii) decompose the kinetic energy into compressible and incompressible components to track energy exchange between vortices and phonons.
Together, these results reveal the VR-RP crossover, delineate the parameter space where it occurs, 
and clarify how compressibility, confinement, and finite temperature govern excitation lifetimes and decay pathways. 
Beyond establishing a clean 3D fingerprint of the JR family, our findings provide a quantitative benchmark for nonequilibrium vortex-phonon dynamics in quantum fluids and open a pathway toward studying device-relevant dissipative mechanisms. 
Finally, we demonstrate that our protocol enables controlled studies of ring-ring interactions and leapfrogging dynamics. 
All results are obtained from  finite-temperature classical-field simulations.

\begin{figure}[t]
\includegraphics[width=1.0\linewidth]{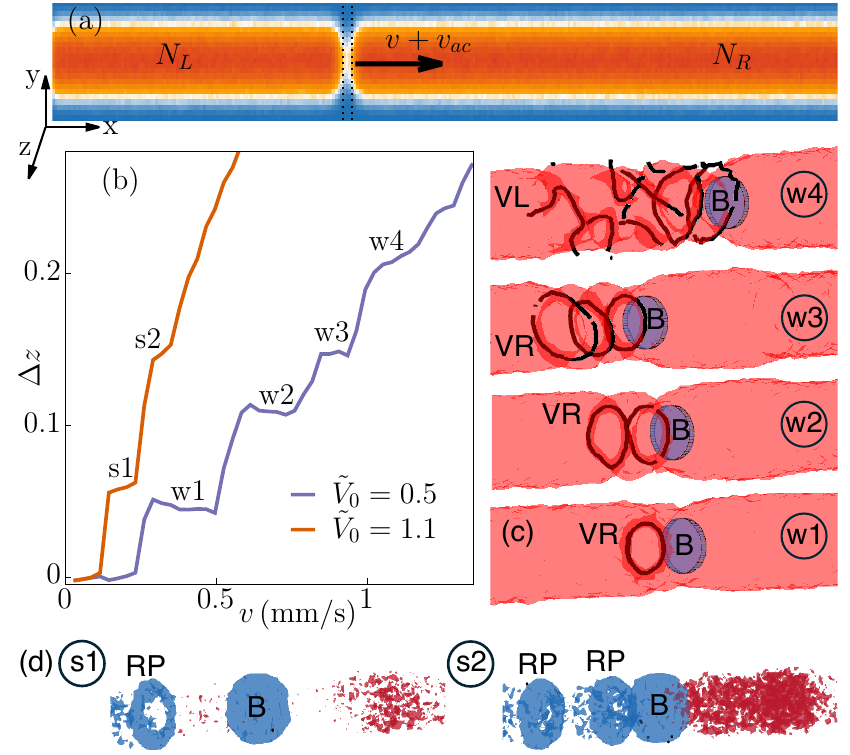}
\caption{On-demand generation of Jones-Roberts excitations in trapped 3D BEC. 
(a) Simulation of the atomic Josephson junction (AJJ), which is created by separating two 3D clouds with a Gaussian barrier of height $V_0$ 
and width $w$ (indicated by two vertical dotted lines). 
$N_L$ ($N_R$) represents the atom number of the left (right) reservoir. 
We use $w=\SI{1.1}{\um}$, and $\tilde{V}_0 \equiv V_0/\mu$ in the range $0.4-1.65$, 
where $\mu$ is the transverse trap-averaged chemical potential. 
dc and ac drives are implemented via the barrier position 
$x(t) = v t + x_1 \sin(2\pi f t)$, 
where $v$ is the dc velocity, $f$ is the ac frequency, 
and $x_1$ is the ac amplitude related with the ac velocity $v_{ac}= 2\pi f  x_1$. 
(b) Velocity-imbalance $(v-\Delta z)$ characteristics for $\tilde{V}_0 =0.5$ and $1.1$.
(c) Semi-transparent isosurface  $n(\br, t)/n_0=0.2$ shows the emission  of one, two, three, and four vortex rings (VRs) at first, second, third, and fourth Shapiro steps, respectively, for $\tilde{V}_0=0.5$.  
The oscillating barrier (blue disc, labeled B) emits both VRs and phonons per drive cycle. 
Extracted vortex polylines overlaid as black tubes; 
closed loops correspond to VRs and open curves to vortex lines (VLs).
(d) At $\tilde{V}_0=1.1$, the barrier nucleates rarefaction pulses (RPs), 
visible as hallow reduced density shells in the background subtracted isosurfaces. 
One and two RPs per drive cycle occur at first and second Shapiro steps, respectively. 
The density ripples (reduction as blue and increase as red) depict phonons. 
}
\label{Fig:system}
\end{figure}

\section*{Simulation method}\label{method}
We consider a weakly interacting Bose-Einstein condensate (BEC) of $^{87}$Rb atoms confined in a 3D tube-like geometry interrupted by a narrow, movable optical barrier that forms an AJJ \cite{Bernhart2024, SinghAJJ2025}. 
The barrier is driven by combined dc and ac motions to induce Shapiro steps \cite{SinghShapiro, Bernhart2024, Del_Pace2024}.  
We simulate the dynamics using a classical-field method within the truncated Wigner approximation \cite{Blakie2008, Polkovnikov2010, Singh2016, Singh2020sound}.  
The system is described by the Hamiltonian
\begin{align} \label{eq:hamil}
\hat{H}_0 &= \int \mathrm{d}{\bf r} \Big[  \frac{\hbar^2}{2m}  \nabla \hat{\psi}^\dagger({\bf r}) \cdot \nabla \hat{\psi}({\bf r})  + V_\mh({\bf r}) \hat{\psi}^\dagger({\bf r})\hat{\psi}({\bf r})  \, \nonumber  \\
&    \quad + \frac{g}{2} \hat{\psi}^\dagger({\bf r})\hat{\psi}^\dagger({\bf r})\hat{\psi}({\bf r})\hat{\psi}({\bf r})\Big],
\end{align}
$\hat{\psi}$ ($\hat{\psi}^\dagger$) is the bosonic annihilation (creation) operator. 
The 3D interaction parameter is given by $g=4\pi a_s \hbar^2/m$, 
where $a_s$ is the $s$-wave scattering length and $m$ is the mass. 
We employ the parameters corresponding to the $^{87}$Rb platform used in recent experiments \cite{Bernhart2024,SinghAJJ2025}, where $a_s= 5.3\, \mathrm{nm}$.
The external confinement is a harmonic potential $V_\mh({\bf r}) = m(\omega_x^2 x^2 + \omega_y^2 y^2 + \omega_z^2 z^2)/2$ with frequencies  $(\omega_x, \, \omega_y, \, \omega_z) = 2\pi \times (1.6,\, 252, \,250)\, \mHz$. 
Within the classical-field approximation, we replace the operators $\hat{\psi}$ in Eq. \ref{eq:hamil} and in the equations of motion by complex numbers $\psi$.  We map real space on a lattice system of  $320 \times 25 \times 25$ sites with the discretization length of $\SI{0.25}{\um}$. 
We sample the initial states $\psi (\br,t=0)$ from a grand canonical ensemble at  temperature $T$ and chemical potential $\mu_0$ 
via a classical Metropolis algorithm. 
To obtain clear snapshots and extended lifetimes of the solitary waves, the temperature is set to $T= 2.2\, \mathrm{nK}$. At higher temperatures, the qualitative behavior persists, but the excitations decay faster \cite{Bernhart2024, SinghAJJ2025}. 
We adjust $\mu_0$ so that the total atom number is $N \simeq 165,000$, 
leading to the maximum density $n_0= \SI{208}{\um}^{-3}$ at the trap center.
This way, the initial ensemble includes quantum and thermal fluctuations. 
Finally, we propagate each initial state using the equations of motion
\begin{equation}\label{eq:eom}
 i \hbar \dot{\psi}(\br, t) = \Bigl(  - \frac{\hbar^2}{2m} \nabla^2 +  V_\mh(\br, t) +  V_\mb(\br, t) + g|\psi|^2 \Bigr) \psi(\br, t),
\end{equation}
which include the barrier potential of the form
\begin{align}
V_\mb({\bf r},t)  = V_0 (t) \exp \Bigl[- \frac{ 2\bigl( x-x_0- x(t) \bigr)^2}{w^2} \Bigr],
\end{align}
with width $w=\SI{1.1}{\um}$ and center $x_0= \SI{28}{\um}$.
$V_0(t)$ is the barrier's strength and $x(t)$ is the time-dependent location.
We first ramp $V_0$ linearly to its desired value over $200\, \mms$ and wait for $50\, \mms$ to equilibrate.
This effectively separates the condensate into two subclouds, 
thus creating an AJJ by suppressing the tunneling at location $x_0$, see Fig. \ref{Fig:system}(a).  
We then drive the barrier using $x(t) = v t + x_1 \sin(2\pi f t)$, 
where $v$ is the dc velocity, $x_1$ is the ac amplitude, and $f$ is the ac frequency \cite{SinghShapiro}. 
We choose $x_1$ in the range $0.125 - \SI{1}{\um}$, $f=90\, \mHz$, and driving time to $3$ cycles.
We use the transverse trap-averaged chemical potential $\mu=2 \mu_0/3$, 
where $\mu_0$ is the chemical potential at the trap center \cite{SinghAJJ2025}. 
We define the normalized barrier height by $\tilde{V}_0 \equiv V_0/\mu$, 
and vary $\tilde{V}_0$ in the range $0.4-1.65$.

 We calculate the local density $n(\br, t)=|\psi(\br, t)|^2$ and the column density $n(x,t)= \sum_{y, z} n(\br, t)$, and express time in units of the drive period $T_f =1/f=11.11\, \mms$. 
The value of the sound velocity, determined from the propagation of a phonon pulse in the column density, 
is $c_s=1.83\, \mathrm{mm/s}$. 
For velocity-imbalance characteristics, we quantify the imbalance $\Delta z = z - z_\mref$ from atom numbers in the left and right reservoir, where $z_\mref$ is taken from a reference cloud without the barrier.
For vortex identification, we employ a plaquette-based phase-circulation criterion and reconstruct 3D vortex filaments to distinguish vortex rings and lines (see Appendix \ref{App:ring}).

To analyze energy exchange between vortex and sound channels, 
we decompose the total kinetic energy into compressible ($E_k^c$ ) and incompressible ($E_k^i$) components \cite{Nore1997,Horng2009, Numasato2010,Griffin2020}. 
We first compute the density-weighted velocity field $\mathbf{w}(\br)= \mathbf{j}(\br)/\sqrt{n(\br)}$, where $\mathbf{j} = (\hbar/m) \Im(\psi^* \nabla \psi)$. The Helmholtz decomposition 
is performed spectrally by projecting the Fourier-transformed velocity field onto its longitudinal $\mathbf{w}_{||}$ and transverse $\mathbf{w}_{\perp}$ parts, with $\nabla \times \mathbf{w}_{||} =0$ and $\nabla \cdot \mathbf{w}_{\perp} =0$, encoding phonons and vorticity channels, respectively.  
The corresponding energy densities are
\begin{align}
E_k^c = \frac{1}{2} \int |\mathbf{w}_{||}|^2 d\br, \qquad  
E_k^i = \frac{1}{2} \int |\mathbf{w}_{\perp}|^2 d\br. 
\end{align}
The resulting $E_k^c(t)$ and $E_k^i(t)$ traces are expressed in units of nanoKelvin per atom after subtracting the baseline value averaged over the initial equilibrium window. 
This approach cleanly separates vortex-associated (incompressible) and phonon-associated (compressible) motion, allowing direct tracking of vortex-phonon energy conversion during creation and decay dynamics.

\section*{Results}
 We first analyze phase locking between the junction and the ac drive via the imbalance response after three drive cycles. 
The $v-\Delta z$ characteristics exhibit clear Shapiro steps for both weak ($\tilde{V}_0=0.5$) 
and strong ($\tilde{V}_0=1.1$) barriers, confirming synchronization over a broad range, see Fig. \ref{Fig:system}(b).
For weak barriers, steps are sharply resolved, 
while for strong barriers the first two remain visible. 
The corresponding critical velocities without ac drive are $v_c \approx 0.58$ and $0.16\, \mathrm{mm/s}$, respectively. 
Each $n$-th step corresponds to the deterministic emission of $n$ nonlinear excitations per cycle.
At $\tilde{V}_0=0.5$, one, two, three, and four coaxial VRs are emitted per cycle at the first, second, third, and fourth steps, respectively [Fig. \ref{Fig:system}(c)].  
The oscillating barrier injects energy near the junction, shedding VRs and phonons 
whenever the instantaneous barrier velocity exceeds the local critical velocity.

  At higher barrier heights, the nature of the nonlinear excitations shifts from VRs to RPs [Fig. \ref{Fig:system}(d)].
For $\tilde{V}_0=1.1$, the first and second steps yield one and two RPs per cycle, respectively,  
while phonon emission appears as red or blue density ripples.
This VR-RP crossover parallels the weak-link to tunneling transition in dc-driven AJJs \cite{SinghAJJ2025}.

In summary, we find that (i) the Shapiro step index determines the number of nonlinear excitations emitted per cycle, 
(ii) the barrier height selects the excitation type (VR or RP), 
and (iii) our phase-circulation-based filament analysis enables unambiguous classification and dynamical tracking of each excitation.

\begin{figure*}[t]
\includegraphics[width=1.0\linewidth]{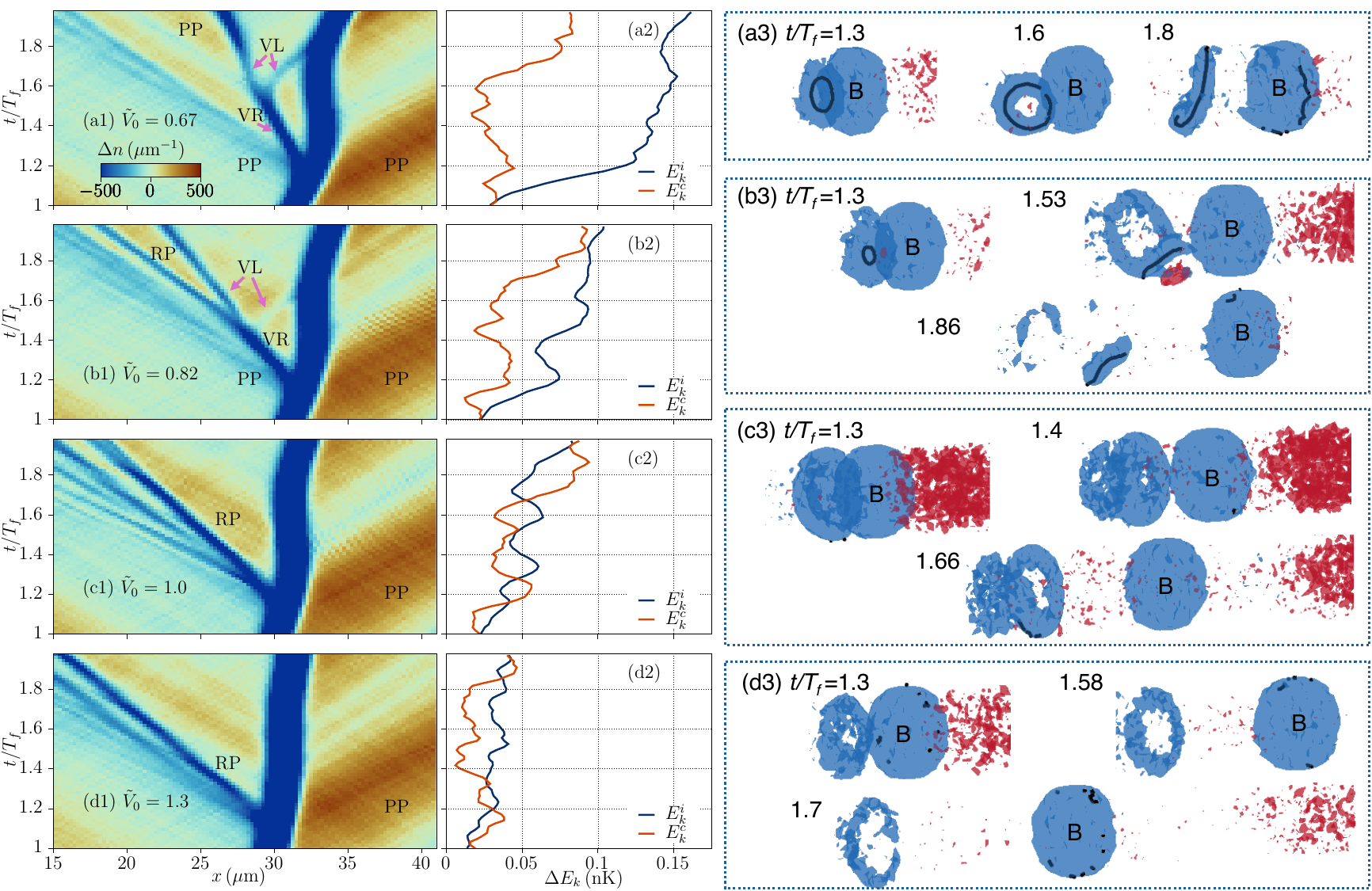}
\caption{Dynamics of JR excitations at the first Shapiro step. 
(a1-d1) Time evolution of the column density $\Delta n (x, t)= n (x, t) - n_0(x)$ at different barrier heights $\tilde{V}_0$, 
where $n_0(x)$ is the density  without the barrier. 
Time is given in units of the drive period $T_f$. 
The oscillating barrier (thick depletion) generates both phononic (linear) and nonlinear excitations, 
with the latter propagating below the sound velocity. 
Density pulses include vortex ring (VR), vortex line (VL), rarefaction pulse (RP), and phononic pulse (PP). 
(a2-d2) Corresponding compressible $E_k^c$ and incompressible $E_k^i$ kinetic energy components.
(a3) For $\tilde{V}_0=0.67$,  the isosurface $\Delta n(\br, t)/n_0=0.2$ shows nucleation of a VR (closed polyline) and phonons (density excess in red) at $t/T_f=1.3$. During  evolution, the VR bends toward the transverse boundary and generally decays into two VLs (open loops) propagating in opposite directions. 
B marks the density reduction at the barrier. 
(b3) For $\tilde{V}_0=0.82$, a small-radius VR is nucleated, which dissipates its energy and momentum into an RP (hollow ring) and a VL (open polyline). Their different characteristic velocities allow them to separate over time.  
(c3, d3) At $\tilde{V}_0=1$ and $1.3$, the barrier mainly emits an RP (hollow ring) and phonons.  
}
\label{Fig:ring-rare}
\end{figure*}

\begin{figure}[t]
\includegraphics[width=1.0\linewidth]{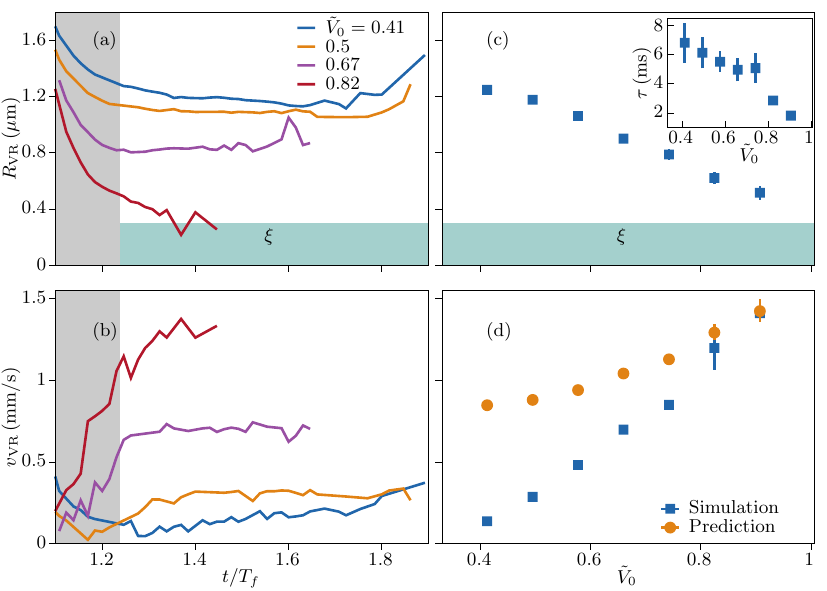}
\caption{Characterization of single vortex rings.
(a, b) Time evolution of the VR radius $R_\mathrm{VR}(t)$ and velocity $v_\mathrm{VR}(t)$ for a single sample in the ensemble at different $\tilde{V}_0$. 
The healing length $\xi \simeq \SI{0.3}{\um}$ (horizontal shaded region) defines the lower bound of $R_\mathrm{VR}$, 
close to which the ring becomes unstable as $R_\mathrm{VR}$ approaches the core radius ($\sim \xi$).
The vertical shaded region indicates the initial build-up time near the barrier.  
(c, d) Averaged $R_\mathrm{VR}$ and $v_\mathrm{VR}$  as a function of $\tilde{V}_0$. 
The inset shows the corresponding ring lifetime $\tau$, averaged over four samples; 
error bars represent the standard deviation. 
In (d), circles represent predictions of the ring velocity $v_\mring$ from Eq. \ref{eq:vring}, using $R_\mathrm{VR}$ values from panel (c); see text.
}
\label{Fig:vel}
\end{figure}

\begin{figure}[t]
\includegraphics[width=1.0\linewidth]{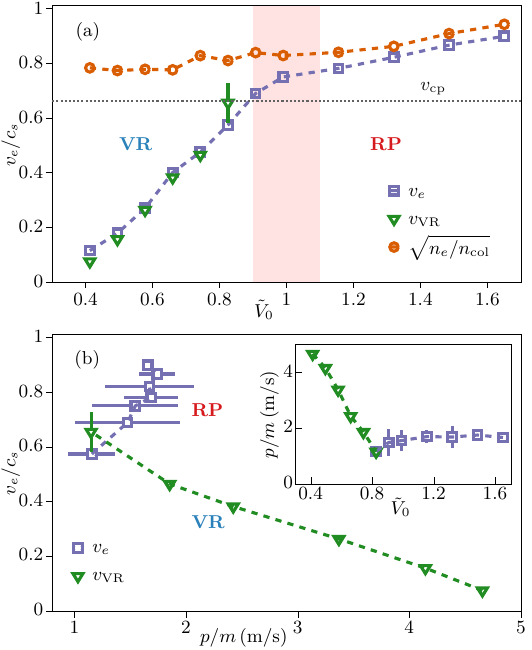}
\caption{Vortex-ring (VR) to rarefaction-pulse (RP) crossover.  
(a) Solitary-wave velocity $v_e$ (squares), normalized by the sound velocity $c_s$, as a function of $\tilde{V}_0$. 
The crossover between VR and RP excitations occurs near $\tilde{V}_0 \sim 1$ (vertical red shaded region).
Triangles show the VR velocity $v_\mVR$ from Fig. \ref{Fig:vel}(d).
The Bogoliubov estimate of the sound velocity, $v_\mathrm{B}/c_s = \sqrt{n_e/n_\mathrm{col}}$ (circles), is based on the reduced density $n_e$ at the dip location, where  $n_\mathrm{col}$ is the column density without the excitation. 
The horizontal dashed line at $v_\mathrm{cp}/c_s  \approx 0.66$ marks the minimum of the energy-momentum dispersion (the cusp) of GPE solutions, 
where the low-velocity (VR) and high-velocity (RP) branches merge \cite{Jones1986}.
(b) $v_e$ and $v_\mVR$ as a function of momentum $p$. The inset shows the VR (triangles) and RP (squares) momenta at varying $\tilde{V}_0$; see text.
Error bars in momentum indicate the uncertainty arising from atom-number estimation within the pulse. 
}
\label{Fig:ex_vel}
\end{figure}
\begin{figure*}
\includegraphics[width=1.0\linewidth]{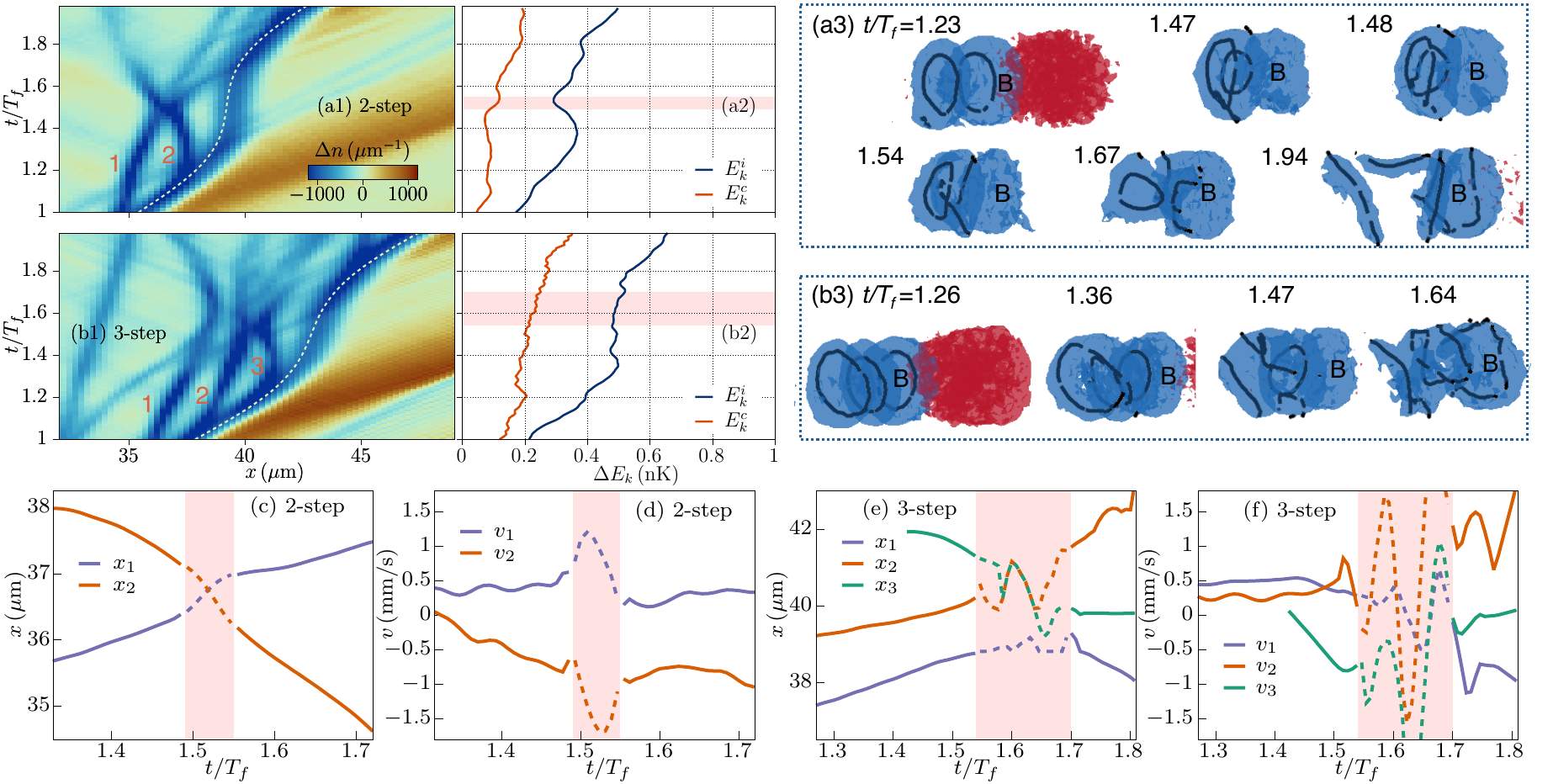}
\caption{Leapfrogging dynamics at higher Shapiro steps for $\tilde{V}_0=0.5$.
(a1, b1) Time evolution of the column density $\Delta n (x, t)$ between cycles $1$ and $2$, showing the nucleation of two and three VRs at the second and third Shapiro steps, respectively. The barrier location is indicated as the dashed line.
When the velocity fields of neighboring VRs overlap, they exhibit leapfrogging dynamics, leading to intricate dynamics for systems with more than two rings. 
(a2, b2) Corresponding compressible $E_k^c$ and incompressible $E_k^i$ components of the kinetic energy per particle.
(a3) Isosurface snapshots of the density difference showing leapfrogging between two sequentially created VRs, which subsequently decay into vortex lines (open loops). 
(b3) Sequential creation and interaction of three VRs at the third Shapiro step.
(c-f) Time evolution of the positions and velocities of the dominant nonlinear density dips at second and third steps, determined after averaging $\Delta n (x, t)$ over $32$ samples.
The shaded region and dotted lines display the scattering time window, where two (c, d) and three (e, f) VRs interact and their subsequent decay dynamics follows. 
Small amplitude oscillations of solid lines and large amplitude oscillations of dashed lines are artifacts of numerical noise and derivatives. 
}
\label{Fig:higherSteps}
\end{figure*}

\subsection*{Excitations at first Shapiro step}
At the first Shapiro step, the oscillating barrier injects energy periodically and drives the local flow above the critical velocity once per cycle.
This produces a reproducible sequence of excitations in each drive period: 
fast phonon pulses propagating near the sound velocity and a slower nonlinear density dip whose character depends on the barrier height, see Fig. \ref{Fig:ring-rare}.
The time evolution of the column density $\Delta n(x, t)$ makes this separation of velocities explicit and  
enables tracking of the dominant nonlinear excitation from its nucleation at the barrier into the bulk of the condensate.

From 3D isosurfaces and vorticity analysis, we identify two regimes.
For moderate barrier heights ($\tilde{V}_0 \approx 0.4-0.8$), a VR is created, propagates through the condensate, and eventually decays.
For higher barriers  ($\tilde{V}_0 \gtrsim 1$), the dominant nonlinear excitation is a RP without phase singularity. 
The column-density evolution reveals broad phonon pulses moving at nearly the sound velocity 
and a slower, more localized dip corresponding to the nonlinear excitation.
The 3D snapshots confirm the topology---closed vortex loops for VRs versus hollow density shells without circulation for RPs.
Cycle-resolved kinetic energy decomposed into incompressible $E_k^i$ and compressible $E_k^c$ components clarifies the underlying energy exchange. 
When a VR is created, $E_k^i$ rises sharply while $E_k^c$ increases only weakly, 
indicating most of the injected energy initially feeds vorticity. 
As the ring decays, energy transfers from $E_k^i$ to $E_k^c$, capturing the sound emission accompanying  breakup. 
In the RP regime, both $E_k^c$ and $E_k^i$ exhibit weak, transient oscillations with no dominant component---consistent with the absence of quantized circulation and the weak propagation of the RP. 
During evolution, the RP continuously radiates energy into sound, visible as small density clouds detaching from the main pulse.
This gradual energy leakage marks the conversion of the localized excitation into dispersive phonons, 
reflecting its coupling to the continuum of sound modes. 
The RP regime thus represents the smooth, high-velocity end of the JR family, 
where vortex-free RPs continuously merge into the phonon branch \cite{Jones1986}.

The decay pathways are governed by confinement and background inhomogeneity. In the VR regime, image-flow attraction tilts the ring toward the transverse boundary; contact with the edge breaks the loop into two dominant vortex lines (VLs), one continuing in the ring's direction and one reflecting toward the barrier. 
This event is accompanied by a clear transfer of energy from $E_k^i$ to $E_k^c$, signaling sound emission. 
This boundary-assisted breakup is the dominant decay channel for barriers up to  $\tilde{V}_0 \sim 0.7$ [Figs. \ref{Fig:ring-rare}(a1-a3)].
For intermediate barriers, smaller ring radii and stronger damping lead to mixed decay channels producing both an RP and a residual VL [Figs. \ref{Fig:ring-rare}(b1-b3)]. 
At higher barriers, the junction operates fully in the RP regime, and no topological breakup occurs; 
the resulting excitation disperses gradually while continuously carrying momentum away from the junction [Figs. \ref{Fig:ring-rare}(c1-d3)]. 
We emphasize that internal ring-core excitations (Kelvin modes) remain negligible in the single-excitation dynamics, confirming that the decay is dominated by boundary interactions rather than intrinsic instabilities.

   We next analyze the geometry and kinematics of the emitted VR.
 The time evolution of the ring radius $R_\mVR (t)$ and velocity $v_\mVR (t)$ for different $\tilde{V}_0$ is shown in Figs. \ref{Fig:vel}(a, c).
 The rapid early-time decrease of $R_\mVR$ reflects circulation accumulation near the barrier before the formation of a well-defined core. 
 Once formed, both $R_\mVR$ and $v_\mVR$ reach a quasi-steady plateau prior to decay. 
As the barrier height increases, the steady-state $R_\mVR$ decreases while $v_\mVR$ increases---consistent with the JR family of low-velocity solitary waves, in which smaller rings travel faster.
This behavior agrees with the analytical expression for a thin-core VR in a uniform condensate \cite{Roberts1971}, 
\begin{align}\label{eq:vring}
v_\mring = \frac{\kappa}{4\pi R} \Bigl[  \ln \Bigl( \frac{8 R}{a} \Bigr)   - 0.615 \Bigr],
\end{align}
where $\kappa= h/m$ is the circulation quantum and $a \simeq \xi$ is the core radius. 
Using simulated $R_\mVR$ as input yields reasonable agreement with observed velocities, 
especially at higher $\tilde{V}_0$, where boundary effects are minimal and the motion approaches the uniform-space prediction [Fig. \ref{Fig:vel}(d)].

 To unify the VR-RP crossover, we extract the solitary-wave velocity $v_e$ of the dominant nonlinear dip from $\Delta n (x, t)$ and compare it with both the tracked ring velocity $v_\mVR$ and a local Bogoliubov estimate $v_\mB/c_s = \sqrt{n_e/n_\mathrm{col}}$, based on the reduced density $n_e$ at the dip center and the density $n_\mathrm{col}$ without the excitation [Fig. \ref{Fig:ex_vel}(a)]. 
 For $\tilde{V}_0 \lesssim 0.9$,  $v_e$ coincides with $v_\mVR$ and follows the decreasing-radius, 
 increasing-velocity trend described above. 
 Around $\tilde{V}_0 \sim 1$, the behavior changes smoothly: $v_e$ increases only weakly 
 and converges toward the RP branch identified from the isosurfaces. 
 This marks the transition from vortex-carrying solitary waves to vorticity-free RPs. 
 Physically, stronger barriers inject shorter-wavelength, more compressive perturbations that fail to sustain a quantized core; 
 energy and geometry then favor a dispersive soliton-like excitation over a topological defect. 

  For a circular (thin-core) VR, the momentum is $p_\mVR = \pi n_0 m \kappa R^2$, 
predicting a quadratic dependence on $R$ \cite{Roberts1971}. 
Using simulated $R_\mVR$ as input yields $p_\mVR$.
For RPs, the total momentum along $x$ is obtained from the continuity equation, 
$p_x \approx m v_\mRP \Delta N$, 
where $\Delta N$ is the atom number reduction within the pulse relative to the unperturbed cloud. 
Computing $\Delta N$ gives $p_\mRP$. 
Both quantities are shown in the inset of Fig. \ref{Fig:ex_vel}(b). 
While $p_\mRP$ varies weakly with $\tilde{V}_0$, 
$p_\mVR$ increases rapidly with decreasing $\tilde{V}_0$. 
In Fig. \ref{Fig:ex_vel}(b), $v_e$ and $v_\mVR$ are plotted versus total momentum $p$, 
revealing two branches consistent with the JR dispersion \cite{Jones1982}: 
the VR branch exhibits increasing velocity with decreasing momentum up to a cusp, 
beyond which the RP branch emerges with increasing velocity approaching the sound velocity.

Together, these results show that (i) the junction at the first Shapiro step emits one well-defined nonlinear carrier per cycle; 
(ii) the carrier evolves continuously from a quantized vortex ring to a vorticity-free rarefaction pulse as $V_0$ increases; and
(iii) the partition of kinetic energy into incompressible and compressible components provides a direct dynamical fingerprint of the excitation's topology.

\subsection*{Dynamics at higher steps}
We now exploit the ability to reproducibly nucleate VRs to probe their nonlinear dynamics at higher Shapiro steps. 
Throughout this section, the barrier height is fixed at $\tilde{V}_0=0.5$, 
and we analyze the behavior at the second and third steps in Fig. \ref{Fig:higherSteps}.


At the second Shapiro step, the barrier ejects two like-signed, coaxial VRs in sequence.
As the rings approach one another, their velocity fields overlap and induce axial flows via Biot-Savart-like induction. 
This mutual coupling generates opposite radial strains:  
the leading ring expands while the trailing ring compresses.
Because the self-induced velocity of a ring decreases with radius, the leading ring slows and the trailing ring accelerates, resulting in a characteristic pass-through or ``leapfrogging" motion. 
In an ideal, unbounded, incompressible fluid this motion would be nearly periodic \cite{Saffman1992, Barenghi2009, Caplan2014, Ikuta2019}.
In our trapped, compressible superfluid, however, it is weakly inelastic: the accelerating, curved filaments radiate sound during each leap, seen as coincident drops in $E_k^i$ and spikes in $E_k^c$. 
While a detailed analysis of internal-core excitations is challenging within the present setup, indications emerge that Kelvin-mode oscillations may be excited during multi-ring interactions. 
Confinement further destabilizes the expanded (slower) ring; upon contacting the Thomas-Fermi (TF), boundary, it fragments into VLs, followed by the decay of the trailing ring.

  At the third Shapiro step, three like-signed rings are emitted sequentially [Figs. \ref{Fig:higherSteps}(b1-b3)].  
The first two initially display the same leapfrogging behavior, but the arrival of the third ring reshapes the induction landscape: its flow slows and expands the second ring, counteracting the compression caused by the first. 
The resulting push-pull competition frustrates clean two-body leapfrogging: three-body induction drives the leading rings toward the TF boundary, where both destabilize and break into VLs. 
The resulting mixture of vortex lines and emitted phonons perturbs the third ring, which briefly accelerates before decaying. 
Corresponding energy traces again show bursts in $E_k^c$ synchronized with declines in $E_k^i$, indicating repeated vortex-to-sound energy transfer during the cascade. 

To quantify these dynamics, we track the positions of the dominant nonlinear density dips from the column-density evolution $\Delta n(x, t)$, averaged over $32$ samples to suppress background and residual excitation effects. Time derivative of these positions yields the instantaneous velocities. 
The corresponding trajectories and velocities for the second and third Shapiro steps are shown in Figs. \ref{Fig:higherSteps}(c-f). In the two-ring case (second step), the locations $x_1(t)$ and $x_2(t)$ exhibit alternating acceleration and deceleration within a well-defined scattering window (shaded region), marking successive leapfrogging events.
The associated velocities $v_1(t)$ and $v_2(t)$ are out of phase, confirming reciprocal induction between the two rings: each expansion of the leading ring coincides with compression and acceleration of the trailing one.
In the three-ring case (third step), the trajectories $x_{1,2,3}(t)$ display more complex coupling. Initially, the first two rings undergo partial leapfrogging, but as the third ring enters the interaction region, mutual three-body induction alters the phase relation---manifesting as irregular velocities changes and enhanced damping. The scattering window delineates the period during which the three rings overlap and exchange momentum before decaying into VL fragments. 
Overall, this quantitative analysis demonstrates that energy and momentum transfer during multi-ring collisions proceeds through repeated, inelastic leapfrogging cycles.


\section*{Discussion and outlook}
A driven atomic Josephson junction provides a compact and tunable platform to generate and study the complete Jones-Roberts (JR) family of solitary waves in three dimensions. 
By tuning the barrier height and drive amplitude, the system continuously spans the spectrum from quantized vortex rings (VRs) to vorticity-free rarefaction pulses (RPs). 
The Shapiro-step phenomenon offers direct control over the number of emitted excitations: the first step generates a single nonlinear excitation, while higher Shapiro steps produce multiple excitations per cycle. This precise control enables reproducible access to distinct dynamical regimes and establishes a new route for deterministic  generation of topological and nontopological solitary waves in superfluids.

The combined analysis of geometry, kinematics, and energetics, reveal how these excitations exchange energy with sound, clarifying  the microscopic origin of phase-slip dissipation and vortex-phonon coupling in driven junctions. Multi-ring emission further enables controlled studies of leapfrogging dynamics, sound-mediated decay, and vortex-line formation---key processes of superfluid turbulence and atomtronic transport.

Future extensions using in-situ imaging could directly visualize the emission and decay of these nonlinear excitations.
Such measurements can be realized using selective imaging techniques \cite{Sunami2022, Sunami2023} capable of reconstructing entire 3D condensate profiles slice by slice. 
Applying this protocol to multicomponent condensates may reveal hybrid vortex-soliton structures and intercomponent drag effects \cite{Ruostekoski2001,KASAMATSU2005}.  
These directions will expand the landscape of nonlinear excitations and advance the design of superfluid quantum circuits.

\section*{ACKNOWLEDGMENTS} 
We acknowledge fruitful discussions with Erik Bernhart and Joachim Brand.
L.M. acknowledges funding by the Deutsche Forschungsgemeinschaft (DFG) in the excellence cluster  `Advanced Imaging of Matter’ - EXC 2056 - project ID 390715994 and by ERDF of the European
Union and `Fonds of the Hamburg Ministry of Science,
Research, Equalities and Districts (BWFGB)’. H.O. acknowledges financial support by the DFG within the SFB OSCAR (project number 277625399).

\bibliography{References}

\clearpage

\appendix
\renewcommand{\thefigure}{A\arabic{figure}}
\renewcommand{\theequation}{A\arabic{equation}}
\setcounter{figure}{0}
\setcounter{equation}{0}


%
%
%
\section{Calculation of vortex filaments}\label{App:ring}
We extract vortex filaments from $\psi(\br, t)$ using a plaquette-phase criterion. 
For each grid face we compute the wrapped phase circulation $\Gamma_\Box = \sum_{j}^4 \Delta \phi_j $, where $\Delta \phi_j = \phi_{j+1}- \phi_j  \in (-\pi, \pi]$.  
Faces with $|\Gamma_\Box| > \pi/2$ are tagged as  pierced by a vortex.
Core positions are then refined to sub-grid accuracy by bilinear intersection of $\Re \psi=0$ and $\Im \psi=0$. 
Nearby intersection points are paired into short segments and merged with a small tolerance; segments are then assembled into continuous polylines by an adjacency graph walk. 
Closed polylines are classified as VRs and fitted to a best-fit plane and circle to obtain ring center, radius and normal. 
Filaments are linked across time frames by nearest-neighbor matching of ring centers to yield trajectories and velocities. 
Open filaments are characterized as vortex lines (VLs).
The analysis is performed within the Thomas-Fermi (TF) radii in $yz$ directions to suppress boundary artefacts. 
This procedure follows established GPE-vortex extraction methods (e.g. Refs. \cite{Zuccher2012,Koplik1993, Villois2016, Leoni2016}) and is numerically robust to thermal noise and finite resolution. 
For visualization we render the calculated polylines as tubes atop semi-transparent density isosurfaces [Fig. \ref{Fig:system}(c)].

%

%

\end{document}